\documentclass[12pt]{article}
\usepackage{graphicx}
\usepackage{rotating}
\renewcommand{\baselinestretch}{1.}
\begin{document}
\begin{center}
{\Large \bf Searching for minimum in dependence of squared
speed-of-sound on collision energy}

\vskip1.0cm

Fu-Hu Liu$^{a,}${\footnote{E-mail: fuhuliu@163.com;
fuhuliu@sxu.edu.cn}}, Li-Na Gao$^{a}$, and Roy A.
Lacey$^{b,}${\footnote{E-mail: Roy.Lacey@Stonybrook.edu}}

{\small\it $^a$Institute of Theoretical Physics, Shanxi
University, Taiyuan, Shanxi 030006, China

$^b$Departments of Chemistry \& Physics, Stony Brook University,
Stony Brook, NY 11794, USA}
\end{center}

\vskip1.0cm

{\bf Abstract:} Experimental results of the rapidity distributions
of negatively charged pions produced in proton-proton ($p$-$p$)
and beryllium-beryllium (Be-Be) collisions at different beam
momentums, measured by the NA61/SHINE Collaboration at the super
proton synchrotron (SPS), are described by a revised
(three-source) Landau hydrodynamic model. The squared
speed-of-sound parameter $c^2_s$ is then extracted from the width
of rapidity distribution. There is a local minimum (knee point)
which indicates a softest point in the equation of state (EoS)
appearing at about 40$A$ GeV/$c$ (or 8.8 GeV) in $c^2_s$
excitation function [the dependence of $c^2_s$ on incident beam
momentum (or center-of-mass energy)]. This knee point should be
related to the searching for the onset of quark deconfinement and
the critical point of quark-gluon plasma (QGP) phase transition.
\\

{\bf Keywords:} Rapidity distribution of negatively charged pions,
speed of sound, revised (three-source) Landau hydrodynamic model,
knee point
\\

PACS: 13.85.-t, 13.85.Ni, 25.75.Dw, 24.10.Nz, 24.10.Pa

\vskip1.0cm

{\section{Introduction}}

Comparing with the relativistic heavy ion collider (RHIC) in USA
[1--3] and the large hadron collider (LHC) in Switzerland [4, 5],
the fixed target experiments performed at the super proton
synchrotron (SPS) in Switzerland [6, 7] present relatively simple
and clean collisions process. The multiplicity in collisions at
the SPS is also low comparing with those at the RHIC and LHC. As
one of the ``first day" measurement quantities, the rapidity
(pseudorapidity) distributions of charged particles are reported
by experimental collaborations [1--7]. These distributions give us
a chance to analyze longitudinal picture of particle productions.
At the same time, based on the rapidity distributions, one can
extract other information such as the penetrating (stopping) power
of projectile (target) nucleus, energy and rapidity losses of
projectile nucleus, energy and particle densities of interacting
region, and so forth.

As a measurement of particle density and mean free path, the
squared speed-of-sound parameter which characterizes partly the
formation of matters in interacting region can be extracted from
the width of Gaussian rapidity distribution described by the
Landau hydrodynamic model [8--18]. Generally, the rapidity spectra
obtained in collisions at the SPS, RHIC, and LHC are not simply
Gaussian distributions. Instead, two Gaussian distributions, one
in the backward (target) region and the other in the forward
(projectile) region, are needed to fit the experimental data [6,
19, 20]. However, the two-Gaussian distribution implies that there
is no central source at midrapidity, which seems unbelievable.
According to the three-source relativistic diffusion model
[21--24], a central source arisen from interactions between
low-momentum gluons in both target and projectile should be
located at midrapidity, a target-like (projectile-like) source
arisen from interactions between valence quarks in the target
(projectile) and low-momentum gluons in the projectile (target)
are expected to appear in the backward (forward) rapidity region.

To extract the squared speed-of-sound parameter so that the
formation of matters can be characterized by an alternative
method, we need the Landau hydrodynamic model [8--10] and its
improved or simplified version [11--18] in the descriptions of
rapidity distributions. In fact, the Landau hydrodynamic model is
not good enough to fit the whole rapidity range. In our recent
work [25, 26], we have revised the Landau hydrodynamic model
[8--18] to a three-source situation. The rapidity distributions
measured in experiments are then described by three Gaussian
distributions. This picture is consistent with the three-source
relativistic diffusion model [21--24].

In this paper, we use the revised (three-source) Landau
hydrodynamic model [25, 26] to describe the rapidity distributions
of negatively charged pions produced in proton-proton ($p$-$p$)
and beryllium-beryllium (Be-Be, exactly $^7$Be-$^9$Be) collisions
at different beam momentums measured by the NA61/SHINE
Collaboration at the SPS. Combining with our previous works [25,
26], we observe the dependence of squared speed-of-sound on
collision energy in a wide range.

The rest of this paper is structured as follows. The model is
shortly described in section 2. Results and discussion are given
in section 3. In section 4, we summarize our main observations and
conclusions.
\\

{\section{The model}}

The revised (three-source) Landau hydrodynamic model used in the
present work can be found in our previous work [25, 26]. To give a
short and clear description of the model, we introduce the main
results of the model in the following.

The Landau hydrodynamic model [8--18] results approximately in a
Gaussian rapidity distribution which does not exactly describe the
experimental data [6, 19, 20]. We have revised the model to three
sources: a central source which locates at midrapidity and covers
the rapidity range as wide as possible, and a target (projectile)
source which locates in the backward (forward) region and revises
the rapidity distribution from the central source [25, 26]. The
experimental rapidity distributions are then described by three
Gaussian functions. And based on the three Gaussian functions, the
experimental pseudorapidity distributions can be described by a
method which makes a distinction between rapidity and
pseudorapidity.

According to the Landau hydrodynamic model [8--18] and our
revision [25, 26], the rapidity distribution, $dN_{ch}/dy$, of
charged particles produced in a given source in high energy
collisions can be described by a Gaussian function
\begin{equation}
\frac{dN_{ch}}{dy}=\frac{N_0}{\sqrt{2\pi}\sigma_X} \exp \bigg[-
\frac{(y-y_X)^2}{2\sigma_X^2} \bigg],
\end{equation}
where $\sigma_X$, $y_X$, and $N_0$ denote the width, peak
position, and normalization constant, respectively; $X=C$, $T$,
and $P$ are for the central, target, and projectile sources,
respectively. The experimental result is in fact a sum weighted by
the three Gaussian functions.

The relation between $\sigma_X$ and the squared speed-of-sound
$c^2_s (X)$ is
\begin{equation}
\sigma_X= \sqrt{\frac{8}{3} \frac{c^2_s (X)}{1-c^4_s (X)} L},
\end{equation}
where $L=\ln(\sqrt{s_{NN}}/2m_p)$ is the logarithmic Lorentz
contraction factor which is independent of $X$, $\sqrt{s_{NN}}$
denotes the center-of-mass energy per pair of nucleons in
nucleus-nucleus collisions and is simplified to $\sqrt{s}$ in
$p$-$p$ collisions, and $m_p$ denotes the rest mass of a proton.
From Eq. (2), $c^2_s (X)$ is expressed by using $\sigma_X$ to be
\begin{equation}
c^2_s (X)= \frac{1}{3\sigma^2_X} \bigg( \sqrt{16L^2+9\sigma^4_X}
-4L \bigg).
\end{equation}
In the above extractions of $\sigma_X$ and $c^2_s (X)$ from the
rapidity distribution, we distinguish accurately the rapidity and
pseudorapidity distributions [25, 26]. In the case of representing
the pseudorapidity distribution in experiment, we also extract
$\sigma_X$ and $c^2_s (X)$ from the hidden rapidity distribution.
Our treatment ensures the method of extraction being concordant.
\\

{\section{Results and discussion}}

\begin{figure}
\hskip-1.0cm \begin{center}
\includegraphics[width=14.0cm]{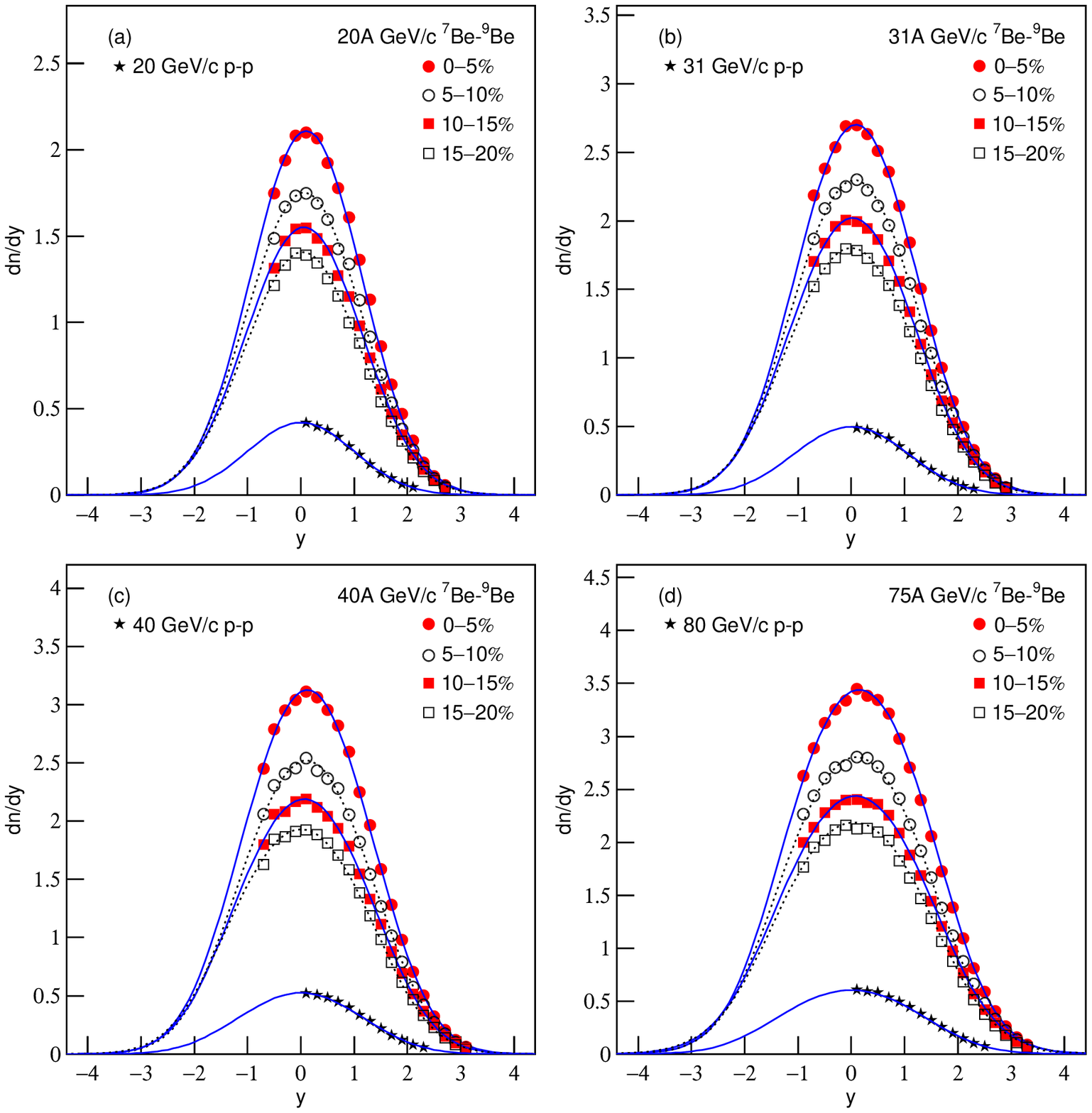}
\vskip.05cm
\includegraphics[width=7.cm]{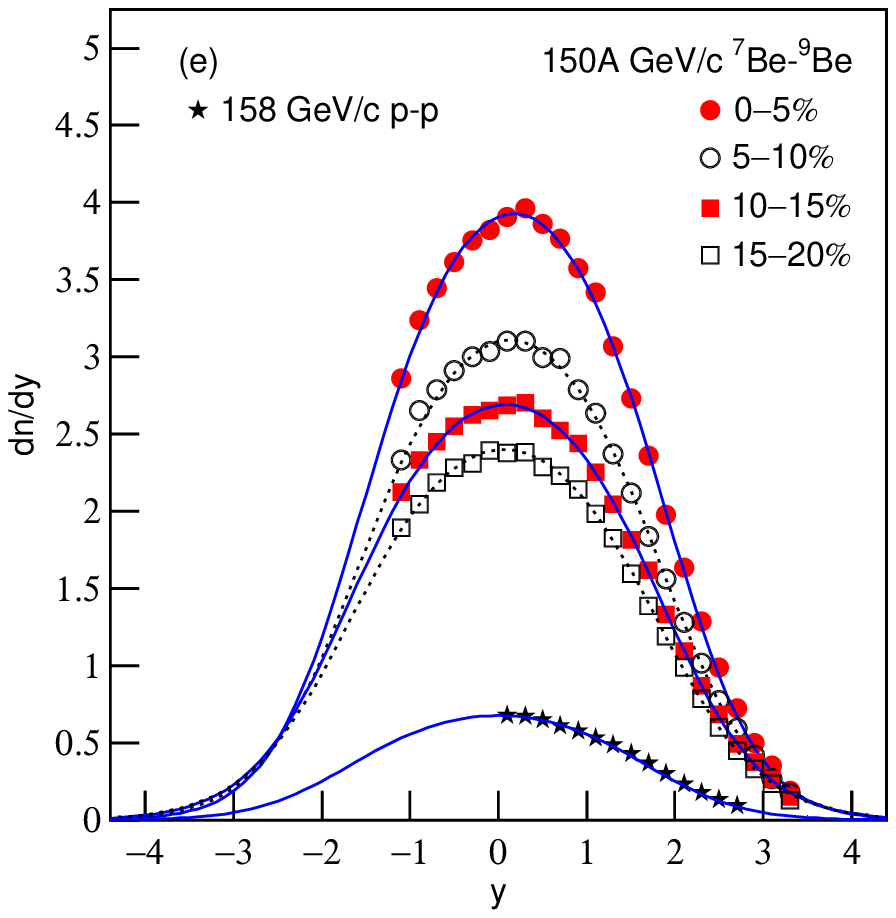}
\end{center}
\vskip-.50cm Figure 1. Rapidity distributions of negatively
charged pions produced in $p$-$p$ and $^7$Be-$^9$Be collisions at
the SPS. Figures 1(a)--1(e) correspond to different beam momentums
marked in the panels. The symbols represent the experimental data
of the NA61/SHINE Collaboration [6, 7] and the curves are our
results fitted by the revised (three-source) Landau hydrodynamic
model.
\end{figure}

Figure 1 presents the rapidity distributions, $dn/dy$, of
negatively charged pions produced in $p$-$p$ and $^7$Be-$^9$Be
collisions at the SPS. For $p$-$p$ collisions, the incident
momentums for Figures 1(a)--1(e) are 20, 31, 40, 80, and 158
GeV/$c$, respectively, which correspond to $\sqrt{s}=6.3$, 7.7,
8.8, 12.3, and 17.3 GeV, respectively. For $^7$Be-$^9$Be
collisions, the incident momentums for Figures 1(a)--1(e) are
$20A$, $31A$, $40A$, $75A$, and $150A$ GeV/$c$, respectively,
which correspond to $\sqrt{s_{NN}}=6.3$, 7.7, 8.8, 11.9, and 16.8
GeV, respectively. The symbols represent the experimental data of
the NA61/SHINE Collaboration [6, 7] and the curves are our results
fitted by the revised (three-source) Landau hydrodynamic model.
Different centrality classes (0--5\%, 5--10\%, 10--15\%, and
15--20\%) for $^7$Be-$^9$Be collisions are presented by different
symbols marked in the panels. The values of fit parameters,
$\sigma_C$, $\sigma_T$ ($=\sigma_P$), $y_C$, rapidity shift
$\Delta y$ ($=y_C-y_T=y_P-y_C$), relative contribution $k_T$ of
the target source ($=k_P$, the relative contribution of the
projectile source), and normalization constant $N_0$, are listed
in Table 1 with the values of $\chi^2$ per degree of freedom
(dof). The last two columns in Table 1 present the values of
$c^2_s(C)$ and $c^2_s(T)$ [$=c^2_s(P)$], which are calculated from
Eq.(3) due to $\sigma_X$. Both $c^2_s(C)$ and $c^2_s(T)$
[$=c^2_s(P)$] are in the units of $c^2$, where $c$ is the speed of
light in vacuum. One can see that the model describes
approximately the experimental data in most cases.

\begin{figure}
\hskip-1.0cm \begin{center}
\includegraphics[width=14.0cm]{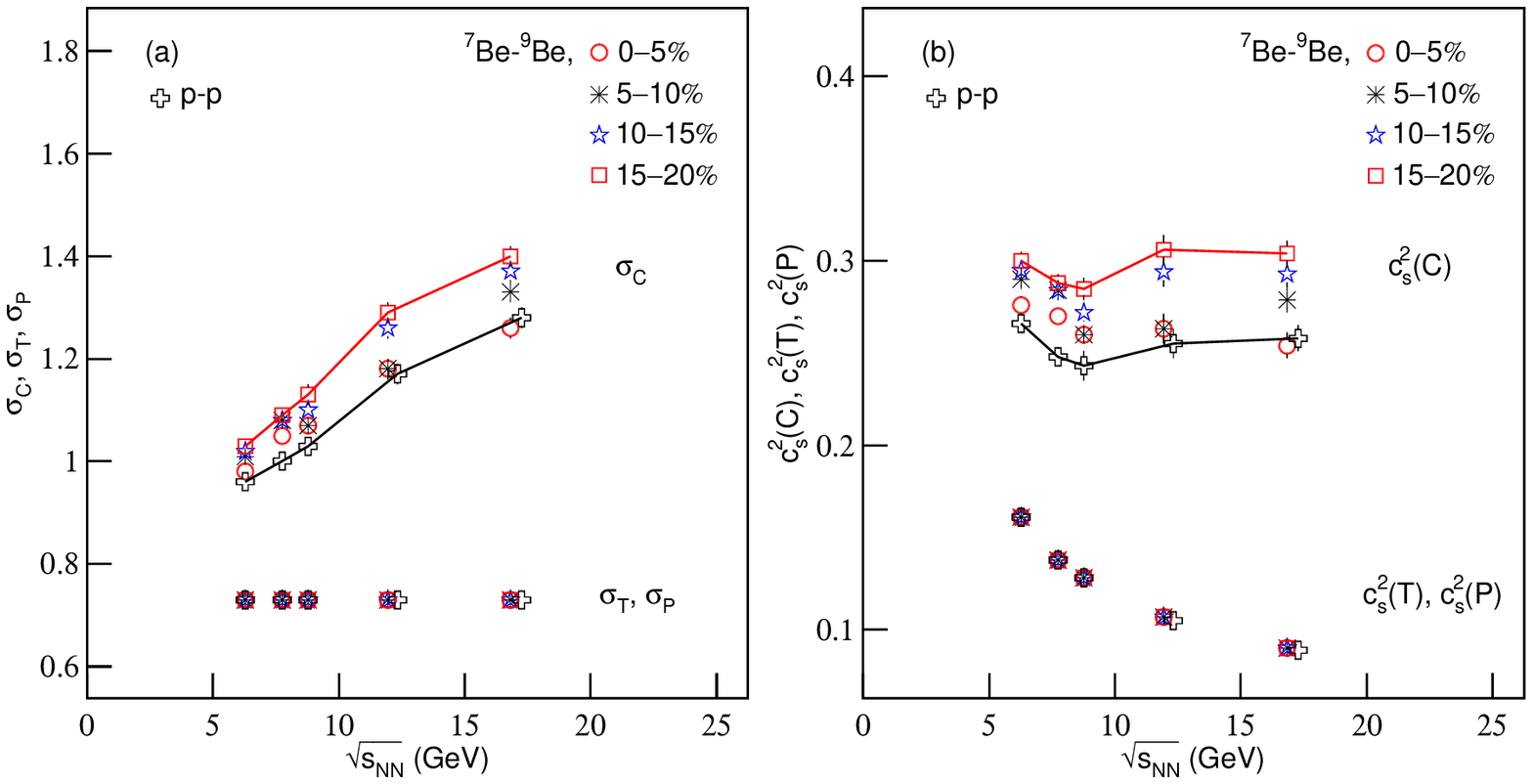}
\end{center}
\vskip.0cm Figure 2. Dependences of (a) $\sigma_X$ and (b)
$c^2_s(X)$ on $\sqrt{s_{NN}}$. The symbols represent the values
listed in Table 1 and the line segments for two cases are given
for guiding the eyes. Different symbols correspond to different
types (centralities) of collisions.
\end{figure}

To see clearly the relation between parameter and energy, Figures
2(a) and 2(b) give the dependences of main free parameter
$\sigma_X$ and extracted parameter $c^2_s(X)$ on $\sqrt{s_{NN}}$,
respectively. The results corresponding to $p$-$p$ collisions and
$^7$Be-$^9$Be collisions with different centrality classes are
displayed by different symbols, which reflect a part of parameter
values listed in Table 1. The line segments for $p$-$p$ collisions
and for $^7$Be-$^9$Be collisions with centrality class 15--20\%
are given for guiding the eyes. One can see that $\sigma_C$
increases with increase of $\sqrt{s_{NN}}$, $\sigma_T$
($\sigma_P$) does not show a change with $\sqrt{s_{NN}}$,
$c^2_s(C)$ shows itself a local minimum (knee point) which
indicates a softest point in the equation of state (EoS) at
$\sqrt{s_{NN}}=8.8$ GeV, and $c^2_s(T)$ [$c^2_s(P)$] decreases
with $\sqrt{s_{NN}}$. The dependences of other parameters ($y_C$,
$\Delta y$, $k_T$, and $N_0$) on energy are not analyzed due to
trivialness.

To study the knee point in detail, Figures 3(a) and 3(b) show
$\sigma_C$ and $c^2_s(C)$ excitation functions [the dependences of
$\sigma_C$ and $c^2_s(C)$ on $\sqrt{s_{NN}}$] respectively, where
the results for $p$-$p$ and Be-Be collisions at lower energies
($\leq 17.3$ GeV) are taken from Figure 2 (Table 1) and the
results for $p$-$p$, proton-antiproton ($p$-$\bar p$),
copper-copper (Cu-Cu), gold-gold (Au-Au), and lead-lead (Pb-Pb)
collisions at higher energies ($\geq 19.6$ GeV) are taken from our
previous works [25, 26]. Different symbols represent different
collisions, which are marked in the panels. The symbols with large
size denote central collisions, and the symbols with small size
denote non-central collisions. The centrality classes for Cu-Cu
collisions are from 0--3\% to 50--55\% [27]; for Au-Au collisions
are from 0--3\% to 45--50\% (maximum 40--45\% at 19.6 GeV) [27];
and for Pb-Pb collisions are from 0--5\% to 20--30\% [28]. The
line segments for $p$-$p$ collisions at SPS and higher energies
are given for guiding the eyes, and the two dotted lines in Figure
3(b) show $c^2_s(C)=1/3$ and 1/2 respectively. One can see that at
higher energies $\sigma_C$ increases with increase of
$\sqrt{s_{NN}}$ and most $c^2_s(C)$ are between 1/3 and 1/2 [a few
$c^2_s(C)$ are beyond 1/2 due to statistical fluctuations].

\begin{figure}
\hskip-1.0cm \begin{center}
\includegraphics[width=14.0cm]{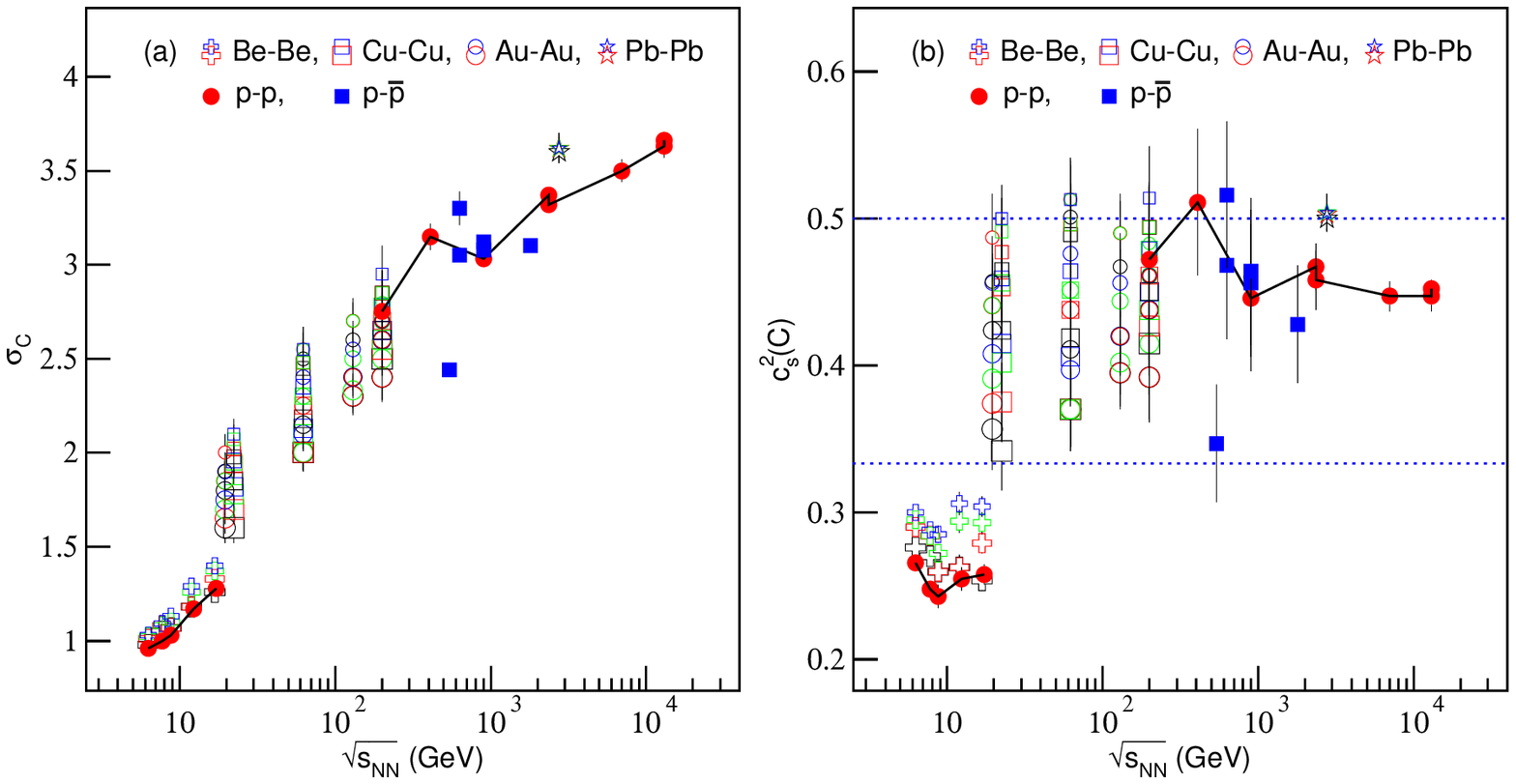}
\end{center}
\vskip.0cm Figure 3. Dependences of (a) $\sigma_C$ and (b)
$c^2_s(C)$ on $\sqrt{s_{NN}}$. The symbols represent the values
listed in Table 1 (for the energy range $\leq 17.3$ GeV), and
taken from our previous work [25, 26] (for the energy range $\geq
19.6$ GeV). The line segments for $p$-$p$ collisions at SPS and
higher energies are given for guiding the eyes, and the two dotted
lines in Figure 3(b) show $c^2_s(C)=1/3$ and 1/2 respectively.
\end{figure}

In particular, one can see some detailed laws in $c^2_s(C)$
excitation function [the dependence of $c^2_s(C)$ on
$\sqrt{s_{NN}}$]: i) In the lower energy range, $c^2_s(C)$
decreases with increase of $\sqrt{s_{NN}}$, then the first knee
point appears at $\sqrt{s_{NN}}=8.8$ GeV, afterwards $c^2_s(C)$
increases with increase of $\sqrt{s_{NN}}$. ii) From 17.3 to 19.6
GeV, $c^2_s(C)$ has a jump from $\leq0.3$ to 1/3--1/2. iii) In the
higher energy range, $c^2_s(C)$ seems not related to
$\sqrt{s_{NN}}$, a saturation appears. The first knee point is a
local minimum which indicates a softest point in the EoS. As the
predecessor of $c^2_s(C)$, $\sigma_C$ increases always with
increase of $\sqrt{s_{NN}}$. We do not observe a softest point in
the EoS in $\sigma_C$ excitation function. These abundant
phenomenons should be related to the searching for the onset of
deconfinement of the quarks and gluons in proton-proton
collisions, and the critical point of phase transition from
hadronic matter to quark-gluon plasma (QGP) in nucleus-nucleus
collisions.

The softest point (8.8 GeV) obtained in the present work is
compatible with the previous works [29, 30] which used the Landau
hydrodynamic model and the ultra-relativistic quantum molecular
dynamics hybrid approach and indicated the softest point locating
in the energy range from 4 to 9 GeV. Other works which study
dependences of ratio of numbers of positive kaons and pions
($K^+/\pi^+$) [20, 31, 32], chemical freeze-out temperature
($T_{ch}$) [31, 32], mean transverse mass minus rest mass
($\langle m_T \rangle -m_0$) [31], and ratio of widths of
experimental negative pion rapidity distribution and Landau
hydrodynamic model prediction [$\sigma_y(\pi^-)/\sigma_y({\rm
hydro})]$ [32] on $\sqrt{s_{NN}}$ show knee point around 7--8 GeV.
A wiggle in the excitation function of a specific reduced
curvature of the net-proton rapidity distribution at midrapidity
is expected in the energy range from 4 to 8 GeV [33, 34]. However,
the searching for the onset of quark deconfinement and the
critical point of QGP phase transition is a complex process [35].
The relation between the softest point and onset of deconfinement
is still an open question. Analyses of more other experimental
data are needed so that they can be confirmed each other.

\begin{figure}
\hskip-1.0cm \begin{center}
\includegraphics[width=14.0cm]{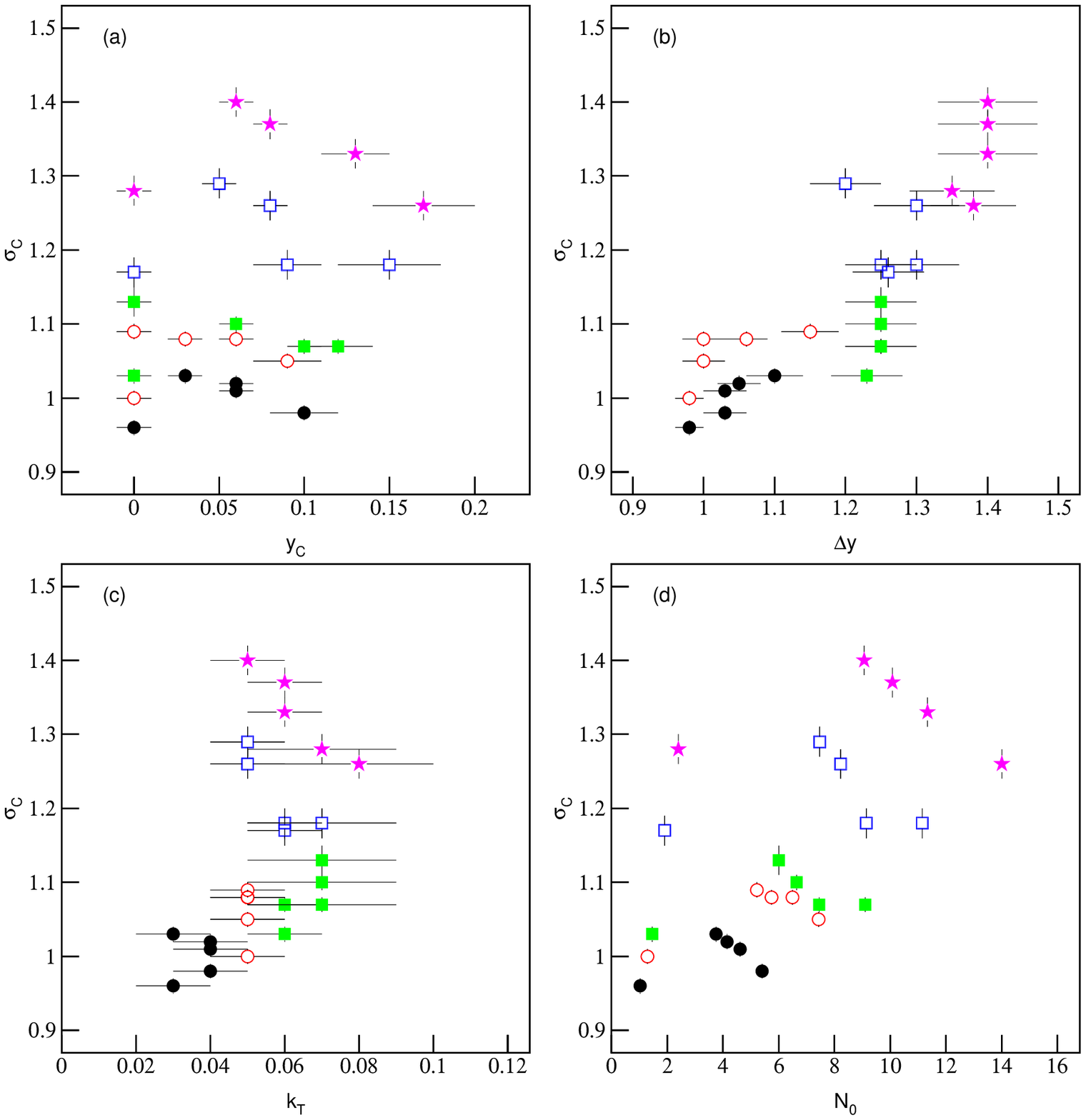}
\end{center}
\vskip.0cm Figure 4. Dependences of $\sigma_C$ on (a) $y_C$, (b)
$\Delta y$, (c) $k_T$, and (d) $N_0$. The symbols represent the
values listed in Table 1, where the closed circles, open circles,
closed squares, open squares, and stars correspond to the
collision energy being 6.3, 7.7, 8.8, 12.3 (11.9), and 17.3 (16.8)
GeV, respectively.
\end{figure}

\begin{figure}
\hskip-1.0cm \begin{center}
\includegraphics[width=14.0cm]{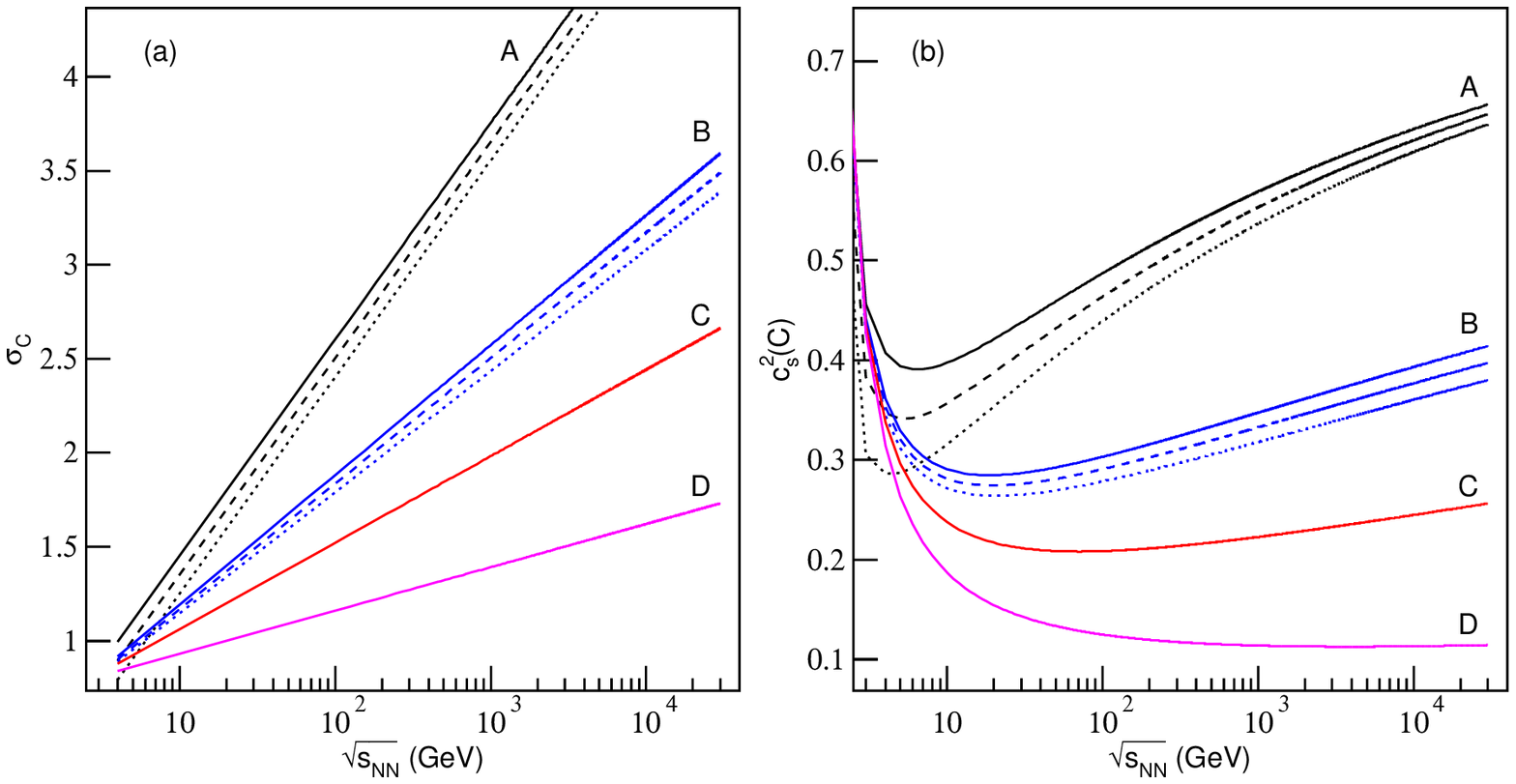}
\end{center}
\vskip.0cm Figure 5. (a) A few examples of linear relations
between $\sigma_C$ and $\ln \sqrt{s_{NN}}$. (b) Different linear
relations between $\sigma_C$ and $\ln \sqrt{s_{NN}}$ corresponding
to different dependences of $c^2_s(C)$ on $\sqrt{s_{NN}}$.
\end{figure}

As for the jump of $c^2_s(C)$ from $\leq0.3$ to 1/3--1/2 when
$\sqrt{s_{NN}}$ increases from 17.3 to 19.6 GeV, we explain it as
the result of changing the mean free path of produced particles,
if the statistical fluctuation is excluded. At the SPS energies,
the interacting system has a small density due to low collision
energy and the produced particles have a large mean free path
which results from a gas-like state and results in a small
$c^2_s$. The situation at the RHIC or LHC energies is opposite,
where the interacting system has a large density due to high
collision energy and the produced particles have a small mean free
path which results from a liquid-like state and results in a large
$c^2_s$. Let $D$ denote the dimensionality of space. According to
Refs. [36, 37], we have the relation of $c^2_s=1/D$ for massless
particles. The particles stay at the gas-like state have a larger
probability to appear in three-dimensional space, which results in
the maximum $c^2_s$ being 1/3 which is the situation at the SPS
energies. The particles stay at the liquid-like state have a
larger probability to appear in two-dimensional space, which
results in the maximum $c^2_s$ being 1/2 which is the situation at
the RHIC and LHC energies.

To see the relations of $\sigma_C$ and other parameters at the SPS
energies which show the softest point in the EoS, Figures 4(a),
4(b), 4(c), and 4(d) present the dependences of $\sigma_C$ on
$y_C$, $\Delta y$, $k_T$, and $N_0$, respectively, where the
closed circles, open circles, closed squares, open squares, and
stars correspond to the collision energy being 6.3, 7.7, 8.8, 12.3
(11.9), and 17.3 (16.8) GeV, respectively, which are taken from
Table 1. One can see that $\sigma_C$ does not show an obvious
dependence on $y_C$ and $k_T$, though $k_T$ shows somehow a
saturation. $\sigma_C$ increases with increases of $\Delta y$ and
$N_0$ due to the latter two increasing with increase of
$\sqrt{s_{NN}}$. The relation between $\sigma_C$ and
$\sqrt{s_{NN}}$ is the most important one among all the relations.
Other relations such as the relations between $\sigma_C$ and
$y_C$, $\Delta y$, $k_T$, as well as $N_0$ are less important. To
present the relations between $c^2_s(C)$ and $y_C$, $\Delta y$,
$k_T$, as well as $N_0$ is trivial due to $c^2_s(C)$ being
calculated from $\sigma_C$.

To study the most important relation between $\sigma_C$ and
$\sqrt{s_{NN}}$ in detail, Figure 5(a) displays a few examples of
linear relations between $\sigma_C$ and $\ln \sqrt{s_{NN}}$ which
reflect approximately the main area of parameter points in Figure
3(a). The solid lines corresponded to cases A, B, C, and D can be
expressed by $\sigma_C = 0.5 \ln \sqrt{s_{NN}} + 0.3$, $\sigma_C =
0.3 \ln \sqrt{s_{NN}} + 0.5$, $\sigma_C = 0.2 \ln \sqrt{s_{NN}} +
0.6$, and $\sigma_C = 0.1 \ln \sqrt{s_{NN}} + 0.7$, respectively.
After conversion by Eq. (3), the four similar lines of $\sigma_C -
\ln \sqrt{s_{NN}}$ show very different relations of $c^2_s(C)- \ln
\sqrt{s_{NN}}$ given correspondingly in Figure 5(b). Cases A and B
show obvious minimums, while cases C and D do not show. Decreasing
the intercept in case A from 0.3 to 0.2 and 0.1 respectively, the
corresponding results presented respectively by the dashed and
dotted lines in Figure 5(a) have a small change, while the results
presented respectively by the dashed and dotted curves in Figure
5(b) have a large change. Decreasing the slope in case B from 0.3
to 0.29 and 0.28 respectively, the corresponding results presented
respectively by the dashed and dotted lines (curves) in Figure
5(a) [5(b)] have a small change. The results presented in Figure 5
show that the minimum in Eq. (3) appearing in some special
conditions.
\\

{\section{Conclusions}}

We summarize here our main observations and conclusions.

(a) The rapidity distributions of negatively charged pions
produced in $p$-$p$ collisions and in $^7$Be-$^9$Be collisions
with different centrality classes at the SPS energies are analyzed
by using the revised (three-source) Landau hydrodynamic model. The
model results are in agreement with the experimental data measured
by the NA61/SHINE Collaboration over an energy range from 6.3 to
17.3 GeV.

(b) The values of squared speed-of-sound parameter are extracted
from the widths of rapidity distributions. For the target
(projectile) source, $c^2_s(T)$ [$c^2_s(P)$] decreases obviously
with increase of $\sqrt{s_{NN}}$ in the considered energy range,
and $\sigma_T$ ($\sigma_P$) does not dependence on
$\sqrt{s_{NN}}$. For the central source, $c^2_s(C)$ shows a
minimum (the softest point in the EoS) at 8.8 GeV when
$\sqrt{s_{NN}}$ increases from 6.3 to 17.3 GeV, and $\sigma_C$
increases obviously with increase of $\sqrt{s_{NN}}$.

(c) Combining with our previous works, one can see that $c^2_s(C)$
is between 1/3 and 1/2 in most cases at higher energies ($\geq
19.6$ GeV). From 17.3 to 19.6 GeV, $c^2_s(C)$ has a jump from
$\leq0.3$ to 1/3--1/2. In the energy range from the minimum SPS
energy to the maximum LHC energy, $\sigma_C$ increases
monotonously with increase of $\sqrt{s_{NN}}$, though some
fluctuations appear in the excitation function.
\\

{\bf Conflict of Interests}

The authors declare that there is no conflict of interests
regarding the publication of this paper.
\\

{\bf Acknowledgments}

The authors would like to thank Dr. Andrzej Wilczek at University
of Silesia in Katowice, Poland and Dr. Emil Aleksander Kaptur at
CERN, Switzerland for supplying us the numerical data of Be-Be
collisions in Figure 1 which might be a little bit different from
that published in ICPAQGP-2015 [7]. This work was supported by the
National Natural Science Foundation of China under Grant No.
11575103 and the US DOE under contract DE-FG02-87ER40331.A008.

\renewcommand{\baselinestretch}{0.8}

\newpage
\begin{sidewaystable}

{\small {Table 1. Values of free parameters, normalization
constants, and $\chi^2$/dof corresponding to the curves in Figure
1. The last two columns are the values of $c^2_s(C)$ and
$c^2_s(T)$ [$=c^2_s(P)$] in the units of $c^2$, where $c$ is the
speed of light in vacuum.
{%
\begin{center}
\begin{tabular}{ccccccccccc}
\hline\hline  Figure & Type & $\sigma_C$ & $\sigma_T$ ($=\sigma_P$) & $y_C$ & $\Delta y$ & $k_T$ ($=k_P$) & $N_0$& $\chi^2$/dof & $c^2_s(C)$ & $c^2_s(T)$ [$=c^2_s(P)$] \\
\hline
Figure 1(a) & $p$-$p$  & $0.96\pm0.01$ & $0.73\pm0.01$ & $0.00\pm0.01$ & $0.98\pm0.02$ & $0.03\pm0.01$ & $1.04\pm0.01$ & 0.284 & $0.266\pm0.005$ & $0.161\pm0.004$ \\
            & 0--5\%   & $0.98\pm0.01$ & $0.73\pm0.01$ & $0.10\pm0.02$ & $1.03\pm0.03$ & $0.04\pm0.01$ & $5.40\pm0.02$ & 6.900 & $0.276\pm0.005$ & $0.161\pm0.004$ \\
            & 5--10\%  & $1.01\pm0.01$ & $0.73\pm0.01$ & $0.06\pm0.01$ & $1.03\pm0.03$ & $0.04\pm0.01$ & $4.62\pm0.01$ & 4.419 & $0.290\pm0.005$ & $0.161\pm0.004$ \\
            & 10--15\% & $1.02\pm0.01$ & $0.73\pm0.01$ & $0.06\pm0.01$ & $1.05\pm0.03$ & $0.04\pm0.01$ & $4.14\pm0.01$ & 3.329 & $0.295\pm0.005$ & $0.161\pm0.004$ \\
            & 15--20\% & $1.03\pm0.01$ & $0.73\pm0.01$ & $0.03\pm0.01$ & $1.10\pm0.04$ & $0.03\pm0.01$ & $3.76\pm0.01$ & 2.494 & $0.300\pm0.005$ & $0.161\pm0.004$ \\
Figure 1(b) & $p$-$p$  & $1.00\pm0.01$ & $0.73\pm0.01$ & $0.00\pm0.01$ & $0.98\pm0.02$ & $0.05\pm0.01$ & $1.30\pm0.01$ & 0.381 & $0.248\pm0.004$ & $0.138\pm0.003$ \\
            & 0--5\%   & $1.05\pm0.01$ & $0.73\pm0.01$ & $0.09\pm0.02$ & $1.00\pm0.03$ & $0.05\pm0.01$ & $7.44\pm0.02$ & 3.090 & $0.270\pm0.004$ & $0.138\pm0.003$ \\
            & 5--10\%  & $1.08\pm0.01$ & $0.73\pm0.01$ & $0.06\pm0.01$ & $1.00\pm0.03$ & $0.05\pm0.01$ & $6.49\pm0.02$ & 4.270 & $0.284\pm0.005$ & $0.138\pm0.003$ \\
            & 10--15\% & $1.08\pm0.01$ & $0.73\pm0.01$ & $0.03\pm0.01$ & $1.06\pm0.03$ & $0.05\pm0.01$ & $5.75\pm0.02$ & 2.951 & $0.284\pm0.005$ & $0.138\pm0.003$ \\
            & 15--20\% & $1.09\pm0.01$ & $0.73\pm0.01$ & $0.00\pm0.01$ & $1.15\pm0.04$ & $0.05\pm0.01$ & $5.22\pm0.02$ & 3.245 & $0.288\pm0.005$ & $0.138\pm0.003$ \\
Figure 1(c) & $p$-$p$  & $1.03\pm0.01$ & $0.73\pm0.01$ & $0.00\pm0.01$ & $1.23\pm0.05$ & $0.06\pm0.01$ & $1.47\pm0.01$ & 1.331 & $0.243\pm0.004$ & $0.128\pm0.004$ \\
            & 0--5\%   & $1.07\pm0.01$ & $0.73\pm0.01$ & $0.12\pm0.02$ & $1.25\pm0.05$ & $0.06\pm0.01$ & $9.11\pm0.02$ & 4.870 & $0.260\pm0.005$ & $0.128\pm0.004$ \\
            & 5--10\%  & $1.07\pm0.01$ & $0.73\pm0.01$ & $0.10\pm0.01$ & $1.25\pm0.05$ & $0.07\pm0.02$ & $7.45\pm0.02$ & 2.841 & $0.260\pm0.005$ & $0.128\pm0.004$ \\
            & 10--15\% & $1.10\pm0.01$ & $0.73\pm0.01$ & $0.06\pm0.01$ & $1.25\pm0.05$ & $0.07\pm0.02$ & $6.64\pm0.02$ & 3.001 & $0.272\pm0.006$ & $0.128\pm0.004$ \\
            & 15--20\% & $1.13\pm0.02$ & $0.73\pm0.01$ & $0.00\pm0.01$ & $1.25\pm0.05$ & $0.07\pm0.02$ & $6.01\pm0.02$ & 5.048 & $0.285\pm0.008$ & $0.128\pm0.004$ \\
Figure 1(d) & $p$-$p$  & $1.17\pm0.02$ & $0.73\pm0.01$ & $0.00\pm0.01$ & $1.26\pm0.05$ & $0.06\pm0.01$ & $1.91\pm0.01$ & 0.528 & $0.255\pm0.008$ & $0.105\pm0.003$ \\
            & 0--5\%   & $1.18\pm0.02$ & $0.73\pm0.01$ & $0.15\pm0.03$ & $1.25\pm0.05$ & $0.06\pm0.01$ &$11.15\pm0.03$ & 4.444 & $0.263\pm0.008$ & $0.107\pm0.003$ \\
            & 5--10\%  & $1.18\pm0.02$ & $0.73\pm0.01$ & $0.09\pm0.02$ & $1.30\pm0.06$ & $0.07\pm0.02$ & $9.15\pm0.02$ &11.861 & $0.263\pm0.008$ & $0.107\pm0.003$ \\
            & 10--15\% & $1.26\pm0.02$ & $0.73\pm0.01$ & $0.08\pm0.01$ & $1.30\pm0.06$ & $0.05\pm0.01$ & $8.23\pm0.02$ & 8.584 & $0.294\pm0.008$ & $0.107\pm0.003$ \\
            & 15--20\% & $1.29\pm0.02$ & $0.73\pm0.01$ & $0.05\pm0.01$ & $1.20\pm0.05$ & $0.05\pm0.01$ & $7.48\pm0.02$ & 7.826 & $0.306\pm0.008$ & $0.107\pm0.003$ \\
Figure 1(e) & $p$-$p$  & $1.28\pm0.02$ & $0.73\pm0.01$ & $0.00\pm0.01$ & $1.35\pm0.06$ & $0.07\pm0.02$ & $2.40\pm0.01$ & 0.912 & $0.258\pm0.007$ & $0.089\pm0.002$ \\
            & 0--5\%   & $1.26\pm0.02$ & $0.73\pm0.01$ & $0.17\pm0.03$ & $1.38\pm0.06$ & $0.08\pm0.02$ &$14.00\pm0.03$ & 8.000 & $0.254\pm0.007$ & $0.090\pm0.002$ \\
            & 5--10\%  & $1.33\pm0.02$ & $0.73\pm0.01$ & $0.13\pm0.02$ & $1.40\pm0.07$ & $0.06\pm0.01$ &$11.33\pm0.03$ & 5.009 & $0.279\pm0.007$ & $0.090\pm0.002$ \\
            & 10--15\% & $1.37\pm0.02$ & $0.73\pm0.01$ & $0.08\pm0.01$ & $1.40\pm0.07$ & $0.06\pm0.01$ &$10.09\pm0.03$ & 2.198 & $0.293\pm0.007$ & $0.090\pm0.002$ \\
            & 15--20\% & $1.40\pm0.02$ & $0.73\pm0.01$ & $0.06\pm0.01$ & $1.40\pm0.07$ & $0.05\pm0.01$ & $9.06\pm0.02$ & 1.567 & $0.304\pm0.007$ & $0.090\pm0.002$ \\
\hline
\end{tabular}%
\end{center}
}} }
\end{sidewaystable}

\end{document}